\documentclass[dvips]{acta}
\usepackage{supertabular,lscape,epsfig}
\usepackage{amssymb}
\usepackage{amsmath}
\usepackage{multirow}
\usepackage{float}
\mathchardef\mhyphen="2D
\usepackage{placeins}
\DeclareSymbolFont{ppa}{OT1}{ppl}{m}{it}
\DeclareMathSymbol{\vv}{\mathalpha}{ppa}{'166}

\SetPages{105}{126}

\SetVol{73}{2023}

\begin{document}
\newcommand\pvalue{\mathop{p\mhyphen {\rm value}}}
\newcommand{\TabApp}[2]{\begin{center}\parbox[t]{#1}{\centerline{
  {\bf Appendix}}
  \vskip2mm
  \centerline{\small {\spaceskip 2pt plus 1pt minus 1pt T a b l e}
  \refstepcounter{table}\thetable}
  \vskip2mm
  \centerline{\footnotesize #2}}
  \vskip3mm
\end{center}}

\newcommand{\TabCapp}[2]{\begin{center}\parbox[t]{#1}{\centerline{
  \small {\spaceskip 2pt plus 1pt minus 1pt T a b l e}
  \refstepcounter{table}\thetable}
  \vskip2mm
  \centerline{\footnotesize #2}}
  \vskip3mm
\end{center}}

\newcommand{\TTabCap}[3]{\begin{center}\parbox[t]{#1}{\centerline{
  \small {\spaceskip 2pt plus 1pt minus 1pt T a b l e}
  \refstepcounter{table}\thetable}
  \vskip2mm
  \centerline{\footnotesize #2}
  \centerline{\footnotesize #3}}
  \vskip1mm
\end{center}}

\newcommand{\MakeTableH}[4]{\begin{table}[H]\TabCap{#2}{#3}
  \begin{center} \TableFont \begin{tabular}{#1} #4 
  \end{tabular}\end{center}\end{table}}

\newcommand{\MakeTableApp}[4]{\begin{table}[p]\TabApp{#2}{#3}
  \begin{center} \TableFont \begin{tabular}{#1} #4 
  \end{tabular}\end{center}\end{table}}

\newcommand{\MakeTableSepp}[4]{\begin{table}[p]\TabCapp{#2}{#3}
  \begin{center} \TableFont \begin{tabular}{#1} #4 
  \end{tabular}\end{center}\end{table}}

\newcommand{\MakeTableee}[4]{\begin{table}[htb]\TabCapp{#2}{#3}
  \begin{center} \TableFont \begin{tabular}{#1} #4
  \end{tabular}\end{center}\end{table}}

\newcommand{\MakeTablee}[5]{\begin{table}[htb]\TTabCap{#2}{#3}{#4}
  \begin{center} \TableFont \begin{tabular}{#1} #5 
  \end{tabular}\end{center}\end{table}}


\newcommand{\MakeTableHH}[4]{\begin{table}[H]\TabCapp{#2}{#3}
  \begin{center} \TableFont \begin{tabular}{#1} #4 
  \end{tabular}\end{center}\end{table}}

\newfont{\bb}{ptmbi8t at 12pt}
\newfont{\bbb}{cmbxti10}
\newfont{\bbbb}{cmbxti10 at 9pt}
\newcommand{\uprule}{\rule{0pt}{2.5ex}}
\newcommand{\douprule}{\rule[-2ex]{0pt}{4.5ex}}
\newcommand{\dorule}{\rule[-2ex]{0pt}{2ex}}
\def\thefootnote{\fnsymbol{footnote}}
\begin{Titlepage}
\Title{The OGLE Collection of Variable Stars.\\
Over 15\,000 ${\pmb \delta}$~Scuti Stars in the Large Magellanic
Cloud\footnote{Based
on observations obtained with the 1.3-m Warsaw telescope at the Las Campanas Observatory of the Carnegie Institution for Science.}}
\Author{I.~~S~o~s~z~y~{\'n}~s~k~i$^1$,~~
P.~~P~i~e~t~r~u~k~o~w~i~c~z$^1$,~~
A.~~U~d~a~l~s~k~i$^1$,~~
J.~~S~k~o~w~r~o~n$^1$,\\
M.\,K.~~S~z~y~m~a~{\'n}~s~k~i$^1$,~~
R.~~P~o~l~e~s~k~i$^1$,~~
D.\,M.~~S~k~o~w~r~o~n$^1$,~~
S.~~K~o~z~{\l}~o~w~s~k~i$^1$,\\
P.~~M~r~{\'o}~z$^1$,~~
P.~~I~w~a~n~e~k$^1$,~~
M.~~W~r~o~n~a$^1$,~~
K.~~U~l~a~c~z~y~k$^{2,1}$,~~
K.~~R~y~b~i~c~k~i$^{3.1}$,\\
M.~~G~r~o~m~a~d~z~k~i$^1$~~
and~~~M.~~M~r~{\'o}~z$^1$
}
{$^1$Astronomical Observatory, University of Warsaw, Al.~Ujazdowskie~4,\\ 00-478~Warszawa, Poland\\
$^2$Department of Physics, University of Warwick, Gibbet Hill Road, Coventry, CV4~7AL,~UK\\
$^3$Department of Particle Physics and Astrophysics, Weizmann Institute of Science, Rehovot 76100, Israel}
\Received{September 26, 2023}
\end{Titlepage}

\Abstract{We present the OGLE collection of $\delta$~Scuti stars in
  the Large Magellanic Cloud and in its foreground. Our dataset
  encompasses a total of 15\,256 objects, constituting the largest
  sample of extragalactic $\delta$~Sct stars published so far. In the
  case of 12 $\delta$~Sct pulsators, we detected additional eclipsing
  or ellipsoidal variations in their light curves. These are the first
  known candidates for binary systems containing $\delta$~Sct
  components beyond the Milky Way. We provide observational parameters
  for all variables, including pulsation periods, mean magnitudes,
  amplitudes, and Fourier coefficients, as well as long-term light
  curves in the {\it I}- and {\it V}-bands collected during the fourth
  phase of the OGLE project.

We construct the period--luminosity (PL) diagram, in which
fundamental-mode and first-overtone $\delta$~Sct stars form two nearly
parallel ridges. The latter ridge is an extension of the PL relation
obeyed by first-overtone classical Cepheids. The slopes of the PL
relations for $\delta$~Sct variables are steeper than those for
classical Cepheids, indicating that the continuous PL relation for
first-overtone $\delta$~Sct variables and Cepheids is non-linear,
exhibiting a break at a period of approximately 0.5~d.

We also report the enhancement of the OGLE collection of Cepheids and
RR~Lyr stars with newly identified and reclassified objects,
including pulsators contained in the recently published Gaia DR3
catalog of variable stars. As a by-product, we estimate the
contamination rate in the Gaia DR3 catalogs of Cepheids and RR~Lyr
variables.}{Stars: variables: delta Scuti -- Stars: oscillations --
  Magellanic Clouds -- Catalogs}

\Section{Introduction}
The fourth phase of the Optical Gravitational Lensing Experiment
(OGLE-IV) has yielded extensive catalogs of variable stars, including
nearly complete samples of Cepheids and RR~Lyr stars in the
Magellanic Clouds (\eg Soszy{\'n}ski \etal 2015a, 2016, 2017,
2019). $\delta$~Scuti variables share the same pulsation mechanism
with Cepheids and RR~Lyr stars, which consequently places them within
the same instability strip in the Hertzsprung-Russell
diagram. Recently, Soszy{\'n}ski \etal (2022) published a collection of
over 2600 $\delta$~Sct pulsators in the Small Magellanic Cloud (SMC)
-- the first ever such catalog covering the entire area of this
galaxy. In this paper, we extend the OGLE Collection of Variable Stars
(OCVS) by about 15\,000 $\delta$~Sct stars carefully selected in the
OGLE-IV photometric database in the Large Magellanic Cloud (LMC).

$\delta$~Sct are mid-A to early-F type pulsating stars that populate
the Cepheid instability strip on or slightly above the main
sequence. They exhibit low-order radial and non-radial pressure modes
with periods below 0.3~d that are self-excited through the
$\kappa$-mechanism. The $\delta$~Sct class includes a mixture of stars
at different evolutionary stages: young stellar objects during their
contraction toward the main sequence, stars with core hydrogen burning
on the main sequence, subgiants evolving off the main sequence, and
Population~II blue stragglers, called SX~Phe stars.

The first $\delta$~Sct stars in the LMC were discovered 20 years
ago. The OGLE-II catalog of RR~Lyr stars in the LMC (Soszy{\'n}ski \etal
2003) was supplemented with 37 short-period pulsating variables, of
which 29 turned out to be actual $\delta$~Sct stars, while the
remaining ones were ultimately classified as Cepheids or RR~Lyr
stars. In turn, Kaluzny and Rucinski (2003) reported the discovery of
eight small-amplitude short-period variables in the LMC open cluster
LW~55 and suggested that seven of them might be $\delta$~Sct stars.

The largest catalogs of $\delta$~Sct stars in the LMC to date were
published based on photometric data collected by the OGLE-III,
SuperMACHO, and EROS-2 surveys. Poleski \etal (2010) identified 2788
candidates for $\delta$~Sct variables in the OGLE-III database,
although more than half of them were marked as uncertain due to their
low luminosity, close to the detection limit of the 1.3-m OGLE
telescope. At the same time, Garg \etal (2010) published a list of
2300 high-amplitude $\delta$~Sct candidates detected by the 4-m
Blanco Telescope operated by the SuperMACHO project. Then, Kim \etal
(2014) reported the discovery of 2481 $\delta$~Sct stars in the
EROS-2 light curve database. These samples were supplemented with
55~$\delta$~Sct variables found by Salinas \etal (2018) in a field
centered on the LMC globular cluster NGC~1846. All these catalogs
together contain about 6600 $\delta$~Sct candidates in the LMC.

In this work, we verify these samples and significantly increase the
number of known $\delta$~Sct variables in the LMC. The remainder of
this paper is organized as follows. In Section~2, we provide details
about the OGLE observations and data reduction. Section~3 outlines the
procedures employed for the identification and classification of
$\delta$~Sct stars in the LMC. In Section~4, we cross-match our
collection of $\delta$~Sct variables with external catalogs of
variable stars. Section~5 presents newly detected Cepheids and RR~Lyr
stars which were also included in the OCVS. As a by-product, we
examine the contamination rates of the recently published Gaia DR3
catalogs of Cepheids (Ripepi \etal 2023) and RR~Lyr stars (Clementini
\etal 2023). In Section~6, we summarize the OGLE collection of
$\delta$~Sct stars in the LMC. The on-sky distribution of $\delta$~Sct
variables in the Magellanic Clouds is presented in Section~7. In
Section~8, we derive the period--luminosity (PL) relations for
$\delta$~Sct stars in the LMC. The subsequent section is devoted to
multimode $\delta$~Sct pulsators. Binary systems containing
$\delta$~Sct components are discussed in Section~10. Finally,
Section~11 provides a summary of our results.
\vspace*{-7pt}
\Section{Observations and Data Reduction}
\vspace*{-3pt}
The OGLE observations were taken with the 1.3-meter Warsaw Telescope
located at Las Campanas Observatory in Chile. The observatory is
operated by the Carnegie Institution for Science. The Warsaw Telescope
is equipped with a mosaic camera consisting of 32 2k$\times$4k CCDs
with about 268 million pixels in total. The field of view of the
OGLE-IV camera is 1.4 square degrees, with a pixel scale of
0\zdot\arcs26. For our research, we utilized photometric data in two
photometric bands obtained within the OGLE-IV project between March
2010 and March 2020. The majority of the observations (typically
around 700 data points per object) were collected by the OGLE-IV
survey using the {\it I}-band filter from the Cousins photometric
system. Additionally, from several to over 300 (typically 120) data
points per star have been secured in the {\it V}-band filter, closely
reproducing the bandpass from the Johnson photometric system.

The OGLE-IV project  regularly observes an area of  765 square degrees
in the Magellanic System region, fully  covering the LMC, SMC, and the
Magellanic Bridge  connecting both galaxies. We  adopted the celestial
meridian of 2\zdot\uph8  as the arbitrary boundary  that separates the
LMC  from the  SMC in  the sky.  In the  LMC region,  the OGLE  survey
monitors the brightness  of around 70 million stars  with {\it I}-band
magnitudes ranging from about 13.0  to 21.5. A detailed description of
the   instrumentation,   photometric   reductions,   and   astrometric
calibrations of the OGLE-IV observations  is provided by Udalski \etal
(2015).
\vspace*{-7pt}
\Section{Search for ${\pmb \delta}$~Sct Stars}
\vspace*{-3pt}
\begin{figure}[b]
\centerline{\includegraphics[width=11.1cm]{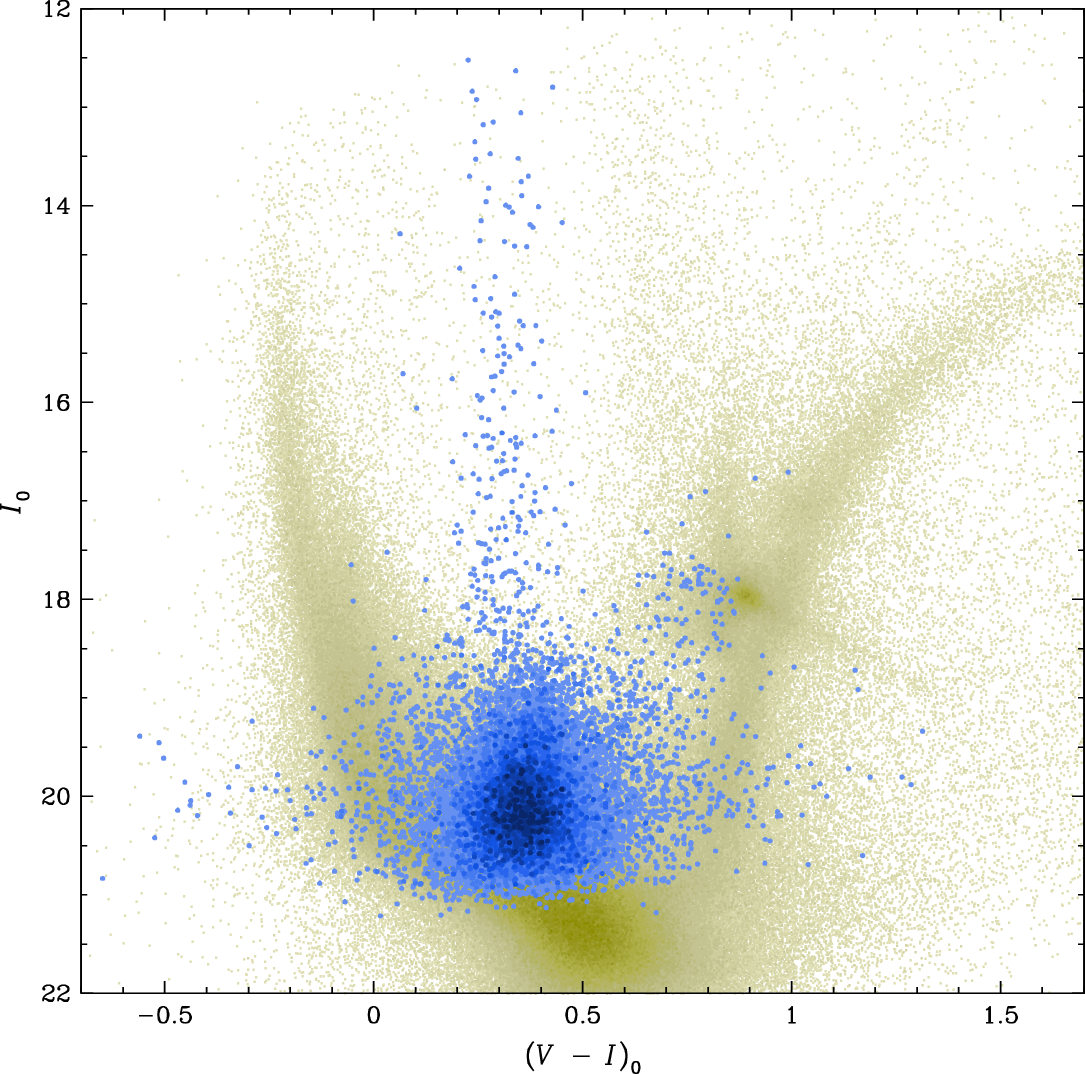}}
\FigCap{$(V-I)_0$ vs. $I_0$ color--magnitude diagram for $\delta$~Sct stars
in the LMC (blue points). The background yellow points show stars from the
field LMC519. Darker colors indicate areas of higher density of the
points. The colors and magnitudes have been corrected for interstellar
extinction using the reddening maps by Skowron \etal (2021).}
\end{figure}
Our search for $\delta$~Sct variables in the LMC followed procedures
similar to those described in Soszy{\'n}ski \etal (2022). Firstly, the
{\it I}-band time series of each star observed by OGLE in the LMC were
passed through the period-search algorithm implemented in the {\sc
Fnpeaks}
code\footnote{\it http://helas.astro.uni.wroc.pl/deliverables.php?lang=en\&active=fnpeaks}. We
explored the frequency range from 0 to 100~cycles per day with a
resolution of $5\times10^{-5}$~cycles per day. For each light curve, we
measured the dominant period and then subtracted it, along with its
first two harmonics, from the data. Then, we repeated the periodicity
search procedure on the residuals, allowing us to measure two
strongest periods for each source.

The  next  stage  of  our  procedure  for  selecting  and  classifying
$\delta$~Sct stars  in the LMC was  based on the visual  inspection of
the  light curves  with a  dominant period  shorter than  0.3~d. After
excluding known eclipsing variables and  RR~Lyr stars, we examined the
{\it I}- and {\it V}-band light curves of over 100~thousand stars with
the largest signal-to-noise ratios of their periods. Although the OGLE
{\it V}-band light  curves consist of a smaller number  of data points
compared to the {\it I}-band time series, this is compensated by lower
noise in the {\it V} filter  for most $\delta$~Sct variables. Based on
the characteristic shapes of the  light curves, we selected an initial
sample of  candidate $\delta$~Sct stars. Furthermore,  we identified a
number of  double-mode pulsators  based on their  high signal-to-noise
secondary periods and period ratios in  the range of 0.75 to 0.81 (see
Section~9).

In  the last  stage,  we  verified our  candidates  by checking  their
positions on  the color--magnitude and  PL diagrams. Fig.~1  shows the
color--magnitude diagram for  $\delta$~Sct stars in the  LMC. The {\it
  I}-band magnitudes and  $(V-I)$ color index have  been corrected for
interstellar extinction  using the  high-resolution reddening  maps by
Skowron  \etal  (2021). During  the  selection  process, most  of  the
sources   with   a   dereddened   color  index   outside   the   range
$0.1<(V-I)_0<0.7$~mag  (corresponding  to  the instability  strip  for
$\delta$~Sct   variables)  have   been   removed   from  our   initial
list. However, approximately 5\% of the  stars in the final version of
our collection  have colors outside  this range because  we considered
that they  may be  genuine $\delta$~Sct  pulsators blended  with other
stars. The  result of our selection  procedure was an initial  list of
approximately 14\,000 $\delta$~Sct stars in the direction of the LMC.

\Section{Comparison with the Literature}
In order to assess the completeness and contamination rate of the OGLE
collection of $\delta$~Sct stars in the LMC, we cross-matched it with
several lists of variable stars, including the catalogs published by
the OGLE-III (Poleski \etal 2010), SuperMACHO (Garg \etal 2010), and
EROS-2 (Kim \etal 2014) projects, the International Variable Star
Index (VSX, Watson \etal 2006), and the Gaia DR3 catalog of
main-sequence pulsators (Gaia Collaboration \etal 2023). We carefully
examined the light curves of $\delta$~Sct candidates that were not
present in the initial version of our collection and supplemented it
with over 1000 objects that we identified as genuine $\delta$~Sct
pulsators. The final OGLE collection contains 4712 stars that were
identified as $\delta$~Sct variables in previously released
catalogs. This means that 10\,544 objects in our sample (69\%) are new
discoveries. Below, we present detailed results of the comparison
between the OGLE collection and other catalogs of $\delta$~Sct stars
in the LMC.

Our collection shares 2309 sources with the OGLE-III catalog of
2788 $\delta$~Sct candidates in the LMC (Poleski \etal 2010). For
the remaining 479 objects (representing about 17\% of the OGLE-III
catalog), we were unable to confirm the classification provided by
Poleski \etal (2010). Several of these stars have been reclassified as
classical Cepheids or RR Lyr variables, several dozen other sources
turned out to be eclipsing or ellipsoidal variables, but the majority
of the rejected stars have an unknown classification. While it is
possible that some of these objects are true $\delta$~Sct stars, we
decided to exclude them from the OCVS in order to maintain the purity
of our sample.  It is worth noting that the vast majority of the
rejected stars were marked as uncertain in the OGLE-III catalog.

The SuperMACHO catalog of high-amplitude $\delta$~Sct stars in the LMC
(Garg \etal 2010) contains 2300 objects\footnote{Garg \etal (2010)
  reported the discovery of 2323 candidates for $\delta$~Sct
  variables, however, 23 of them have been duplicated in the
  SuperMACHO catalog.}. The SuperMACHO project (Rest \etal 2005) was
an optical survey of the LMC conducted with the 4-m Blanco telescope
at the Cerro Tololo InterAmerican Observatory in Chile. The SuperMACHO
photometry is deeper than the OGLE photometry obtained with the
1.3-meter Warsaw telescope, which is the main reason why 308
$\delta$~Sct stars discovered by Garg \etal (2010) are missing in our
collection. In turn, we confirm the classification of 1992 brighter
variables, which indicates a high level of purity of the SuperMACHO
catalog.

The EROS-2 catalog (Kim \etal 2014) comprises 117\,234 automatically
classified variable stars in the LMC, of which 2481 are categorized
as $\delta$~Sct candidates. It is worth noting that the list published
by Kim \etal (2014) contains in total 150\,115 candidates for variable
stars, however objects fainter than $B_E=20$~mag and those with low
signal-to-noise ratios (${\rm S/N}<20$) of their periods were
considered false positives and thus excluded from the official EROS-2
catalog. Following this approach, we also excluded these faint and
low-signal-to-noise EROS-2 sources from our cross-match. Our
collection includes 1376 out of 2481 objects classified by Kim \etal
(2014) as $\delta$~Sct stars, which represents about 56\% of the
EROS-2 sample. Among the missing stars, we found, over 300 eclipsing
binary systems, more than 50 RR~Lyr variables, some irregular
variables, and constant stars.

Comparison of our $\delta$~Sct sample with the VSX catalog (Watson
\etal 2006) revealed 30 common objects, all of which are brighter than
$I=16.5$~mag, indicating they are foreground stars. The vast majority
of these variables were discovered by the All-Sky Automated Survey for
Supernovae (ASAS-SN, Jayasinghe \etal 2019, Christy \etal 2023). In
the catalog of main-sequence pulsators published as part of the Gaia
DR3 (Gaia Collaboration \etal 2023), we identified 49 $\delta$~Sct
stars that overlap with our sample. All of these variables also belong
to the halo of the Milky Way. Additionally, our collection contains 71
variables classified in the Gaia DR3 catalog as RR~Lyr stars,
eclipsing binaries, or short-timescale variables.

\Section{New Cepheids and RR Lyr Stars. Comparison to the Gaia DR3 Catalog}
Our search for $\delta$~Sct variables in the Magellanic System and the
cross-match of the OGLE databases with the Gaia DR3 catalog of
variable stars (Clementini \etal 2023, Ripepi \etal 2023) allowed us
to expand the OGLE collection of Cepheids and RR~Lyr stars (Soszy{\'n}ski
\etal 2015a, 2016, 2017, 2019). Population~I $\delta$~Sct stars and
classical Cepheids form a continuous distribution, therefore we
adopted a boundary pulsation period that separates these two types of
variables. As it was reasoned in Soszy{\'n}ski \etal (2022), we applied a
maximum period of 0.3~d for the fundamental mode and 0.23~d for the
first-overtone mode in $\delta$~Sct stars. Population~I pulsators with
longer periods are categorized as classical Cepheids in the OCVS.

The adoption of these strict criteria resulted in the reclassification
of several pulsating stars already present in the OCVS, although this
change is purely formal. Two multimode variables with the
first-overtone periods shorter than 0.23~d were moved from the list of
classical Cepheids to the collection of $\delta$~Sct stars. On the
other hand, four stars classified by Poleski \etal (2010) as
$\delta$~Sct stars have pulsation periods that, according to our new
criteria, place them among Cepheids, so they were included in the OGLE
collection of classical Cepheids in the LMC (Soszy{\'n}ski \etal
2015ab). In addition, the classification of six pulsators with periods
above 0.2~d has been changed from $\delta$~Sct to first-overtone
RR~Lyr (RRc) stars. In these cases, we mainly relied on their position
in the PL diagram, as they fall on the relation for RRc stars, \ie
below the PL sequence for delta Scuti stars. Table~1 contains all the
reclassified classical pulsators in the LMC that have been moved
between the OGLE catalogs of Cepheids, RR~Lyr stars, and $\delta$~Sct
stars.

\renewcommand{\arraystretch}{1.05}
\MakeTableee{l@{\hspace{12pt}}
l@{\hspace{12pt}}   
c@{\hspace{8pt}}
c@{\hspace{2pt}}}{12.5cm}
{Reclassified variable stars from the OCVS}
{\hline
\noalign{\vskip3pt}
\multicolumn{1}{c}{Old identifier} & \multicolumn{1}{c}{New identifier} & \multicolumn{1}{c}{New} & \multicolumn{1}{c}{Subtype} \\
& & \multicolumn{1}{c}{classification} & \\
\noalign{\vskip3pt}
\hline
\noalign{\vskip3pt}
OGLE-LMC-CEP-3367  & OGLE-LMC-DSCT-07569  & $\delta$~Sct star & 1O/2O \\
OGLE-LMC-CEP-3374  & OGLE-LMC-DSCT-11916  & $\delta$~Sct star & 1O/2O/3O \\
OGLE-LMC-DSCT-0394 & OGLE-LMC-RRLYR-41407 & RR Lyr star      & RRc \\
OGLE-LMC-DSCT-0434 & OGLE-LMC-CEP-4716    & Classical Cepheid & 1O \\
OGLE-LMC-DSCT-0662 & OGLE-LMC-CEP-4717    & Classical Cepheid & F/1O \\
OGLE-LMC-DSCT-0765 & OGLE-LMC-RRLYR-41427 & RR Lyr star      & RRc \\
OGLE-LMC-DSCT-0927 & OGLE-LMC-CEP-4718    & Classical Cepheid & F/1O/2O \\
OGLE-LMC-DSCT-1305 & OGLE-LMC-CEP-4720    & Classical Cepheid & 1O \\
OGLE-LMC-DSCT-1428 & OGLE-LMC-RRLYR-41452 & RR Lyr star      & RRc \\
OGLE-LMC-DSCT-1709 & OGLE-LMC-RRLYR-41461 & RR Lyr star      & RRc \\
OGLE-LMC-DSCT-1955 & OGLE-LMC-RRLYR-41467 & RR Lyr star      & RRc \\
OGLE-LMC-DSCT-2716 & OGLE-LMC-RRLYR-41514 & RR Lyr star      & RRc \\
\noalign{\vskip3pt}
\hline}

Our collection of variable stars has also been enriched with
additional Cepheids and RR~Lyr stars identified as by-products of the
search for $\delta$~Sct pulsators, as well as through cross-matching
with the Gaia DR3 catalog of variable stars (Clementini \etal 2023,
Ripepi \etal 2023). The number of classical Cepheids in the LMC
increased by five objects (including a very rare case of a double-mode
pulsator with the second- and third-overtone modes simultaneously
exited), type~II Cepheids by one, anomalous Cepheids by two, and
RR~Lyr stars by 355 previously overlooked variables. These stars were
omitted in the earlier editions of the OCVS due to a small number of
measurement points in the OGLE light curves, or noisy photometry, or
symmetric light curves, or pulsation periods close to 1/2~d, which led
to an erroneous measurement of the period due to a daily alias. This
update resulted in an increase of the OGLE sample of Cepheids in the
LMC by less than 0.2\%, while the list of RR~Lyr stars grew by less
than 1\%, confirming the high completeness of the OCVS in the
Magellanic Clouds.

We also utilized the Gaia DR3 catalog to validate the completeness of
the OGLE collection of classical pulsating stars in the Galactic bulge
and disk (Udalski \etal 2018, Pietrukowicz \etal 2020, Soszy{\'n}ski \etal
2020, 2021). Firstly, we cross-matched the Gaia catalog with the
OCVS. Then, we extracted and carefully examined the OGLE light curves
of candidate pulsators that were not previously included in our
collection. As a result, the OCVS was supplemented with 108 Galactic
Cepheids of all types (representing 2.9\% of the previously published
sample), 1848 RR~Lyr stars (2.4\%), and 94 $\delta$~Sct stars (0.4\%).

As a by-product of our analysis, we examined the contamination rates of
the Gaia DR3 catalogs of Cepheids and RR~Lyr stars. The final version
of the Gaia catalog contains 15\,006 candidates for classical, type~II,
and anomalous Cepheids in the Milky Way, Magellanic Clouds, M31, and
M33 (Ripepi \etal 2023). The OGLE photometric databases provide
time-series data for 12\,139 of these stars. We confirm that the vast
majority of them, over 97\%, are indeed Cepheids. Among the remaining
stars, we identified more than 50 eclipsing and ellipsoidal variables,
over 50 objects were classified by the OGLE team as RR~Lyr or
$\delta$~Sct stars, we also identified some long-period variables,
spotted variables, and other types of variable stars. Furthermore, the
detailed division into classical and type II Cepheids agrees well in
the Gaia DR3 catalog and OCVS. The only exception are anomalous
Cepheids, as over 40\% of the stars categorized as anomalous Cepheids
in the Gaia DR3 catalog have a different classification in the OGLE
collection.

The Gaia DR3 catalog contains 270\,905 candidates for RR~Lyr stars
(Clementini \etal 2023), out of which 148\,401 are observed by the OGLE
survey. We confirmed the Gaia classification for 104\,346
(approximately 70\%) of these stars. Among the remaining $\approx44\,000$
objects, we discovered around 300 eclipsing variables and some other
types of variable stars. However, the vast majority of the Gaia RR~Lyr
candidates in this group show no periodic variability at all. We
conducted a visual inspection of both the OGLE {\it I}-band and Gaia
{\it G}-band light curves of these misclassified stars and noticed
that most of them are faint ($G>20$~mag), close to the detection limit
of the Gaia telescope. The Gaia time-series photometry for these
objects is usually quite noisy, and the periods provided in the Gaia
DR3 catalog appear to be random measurements fitted to outlier data
points in the light curves. Further analysis revealed a pronounced
correlation between the contamination from non-RR~Lyr sources and the
brightness of stars in the Gaia DR3 catalog. The contamination rate
equals approximately 1\% for objects brighter than $G=19$~mag, it
increases to 21\% for sources with mean {\it G}-band magnitudes
ranging from 19 to 20~mag, and escalates to nearly 90\% for stars
fainter than $G=20$~mag.
\vspace*{-9pt}
\Section{The OGLE Collection of ${\pmb \delta}$~Sct Stars in the LMC}
\vspace*{-3pt}
The definitive version of our collection comprises 15\,256 $\delta$~Sct
variables found in the OGLE fields toward the LMC. Approximately
15\,000 of these stars belong to the LMC, while the remaining objects
are part of the Milky Way's halo. The majority of $\delta$~Sct stars
in our sample exhibit radial pulsation in either the fundamental or
first-overtone mode, as indicated by relatively large amplitudes of
their light curves and position in the PL diagram (see
Section~8). However, we refrain from providing the presumed pulsation
modes of our variables due to the challenges associated with their
identification in specific cases. Instead, we have divided our
$\delta$~Sct sample into single mode and multimode pulsators. The
latter category encompasses 639 stars (about 4\% of the entire
catalog) that feature substantial amplitudes of their secondary or
tertiary pulsation modes.

The list of our $\delta$~Sct stars together with their basic
parameters (equatorial coordinates, intensity-averaged mean magnitudes
in the {\it I} and {\it V} bands, up to three pulsation periods,
amplitudes, epochs of the maximum light, and Fourier coefficients), as
well as OGLE-IV time-series photometry, can be accessed through the
OGLE Internet Archive:
\vskip5pt
\begin{center}
{\it https://ogle.astrouw.edu.pl $\rightarrow$ OGLE Collection of Variable Stars}\\
{\it https://www.astrouw.edu.pl/ogle/ogle4/OCVS/lmc/dsct/}\\
\end{center}
\vskip5pt

For the $\delta$~Sct candidates published in the OGLE-III catalog of
variable stars (Poleski \etal 2010), we maintained their designations
in the format of OGLE-LMC-DSCT-NNNNN (where NNNNN represents a
consecutive number), while only extending the number of digits in the
designation from four to five. The newly added $\delta$~Sct stars have
been organized by their right ascension and given designations from
OGLE-LMC-DSCT-02789 to OGLE-LMC-DSCT-15735.
\vskip5pt

The pulsation periods, along with their uncertainties, were computed
using the {\sc Tatry} code (Schwarzenberg-Czerny 1996) based on the
OGLE-IV light curves obtained between 2010 and 2020. To expand the
temporal coverage of the photometric data from 10 to even over 20
years, the OGLE-IV light curves can be merged with the OGLE-III and
OGLE-II time series provided by Poleski \etal (2010). However, it is
crucial to consider the possibility of zero-point offsets between
these datasets for individual stars.

\begin{figure}[p]
\includegraphics[width=12.8cm]{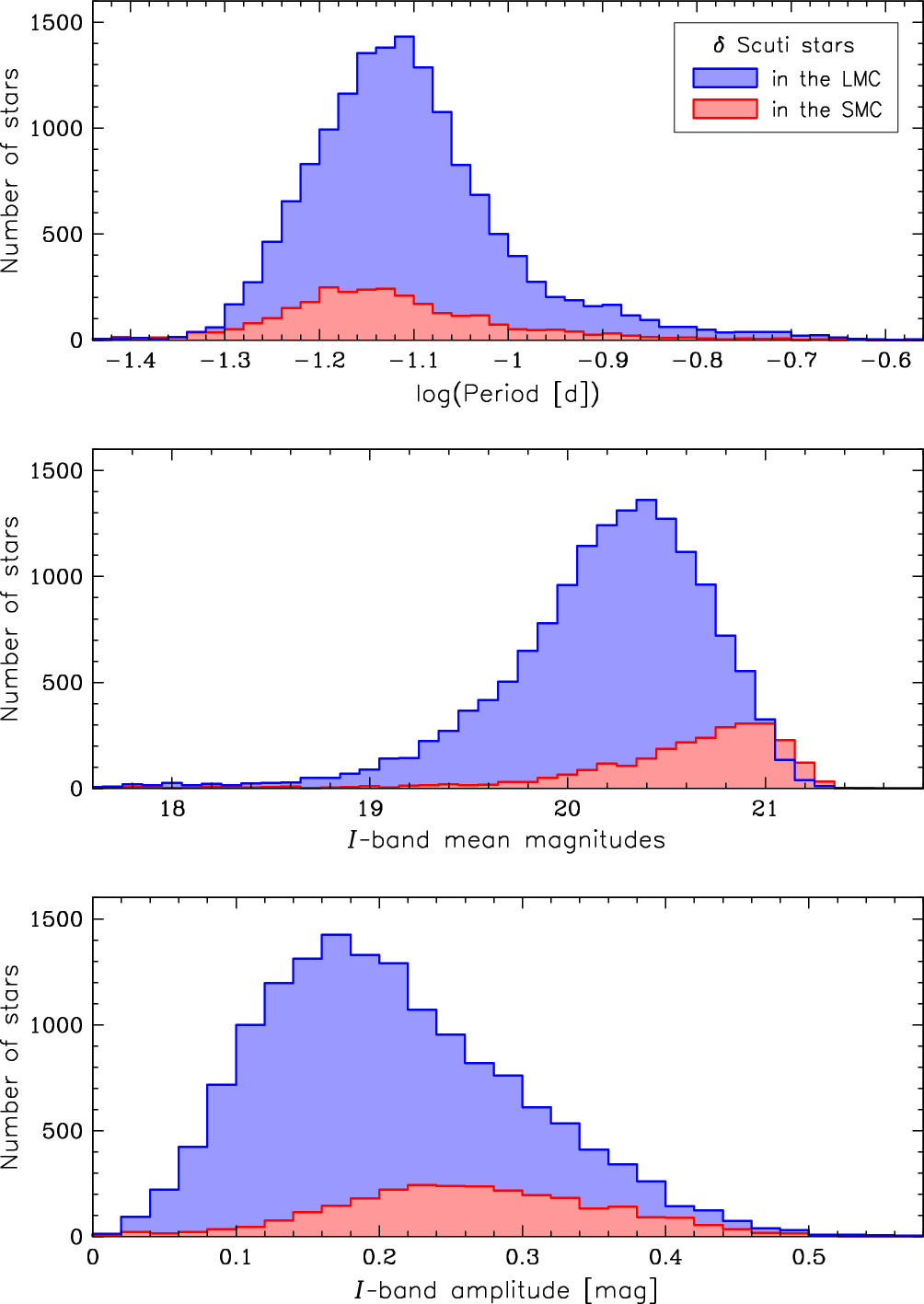}
\vspace*{2pt}
\FigCap{Distributions of dominant pulsation periods ({\it upper panel}),
{\it I}-band mean magnitudes ({\it middle panel}), and {\it I}-band
peak-to-peak amplitudes ({\it lower panel}) of 15\,256 $\delta$~Sct stars in
the LMC (blue histograms) and 2810 $\delta$~Sct stars in the SMC (red
histograms).}
\end{figure}

Fig.~2 illustrates the distributions of pulsation periods (upper
panel), apparent {\it I}-band mean magnitudes (middle panel), and {\it
  I}-band peak-to-peak amplitudes (lower panel) of $\delta$~Sct stars
in the LMC and SMC (Soszy{\'n}ski \etal 2022). The shapes of these
histograms reflect both the intrinsic characteristics of the
$\delta$~Sct population in the Magellanic Clouds and the limitations
of the OGLE photometry. The faintest objects in our collection of
$\delta$~Sct stars in the LMC have mean brightness of about
$I=21.3$~mag, but the number of variables in our sample starts to
decline beyond a luminosity of $I=20.5$ mag. It can be attributed to
the pronounced correlation between the amplitude detection limits and
the observed magnitudes of pulsating stars. For example, for variables
with the mean brightness around $I=20$~mag, the smallest detectable
amplitudes are approximately 0.1~mag, while for stars with $I=21$~mag,
the amplitude detection limit increases to about 0.2~mag.

Keeping in mind these luminosity and amplitude detection limits of the
OGLE survey, we evaluated the completeness of our collection by
considering stars that were identified twice within overlapping
regions of adjacent fields. In the final iteration of our catalog,
each $\delta$~Sct pulsator is uniquely represented by a single entry
from the OGLE database, typically favoring the one with a greater
number of data points in its light curve. We retrospectively checked
that 1162 $\delta$~Sct stars in our collection are located in the
overlapping parts of neighboring OGLE-IV fields, implying that we
could potentially detect 2324 objects from this group. In practice, we
independently confirmed the classification of both components in 624
such pairs, whereas in 538 cases, only one component of the pair was
identified. Consequently, this leads to the catalog completeness level
of approximately 70\%. Once again, we emphasize that this value
applies to $\delta$~Sct stars in the LMC, whose luminosities and
amplitudes are sufficiently large to be detectable with the OGLE
photometry.

\begin{figure}[p]
\centerline{\includegraphics[width=12cm]{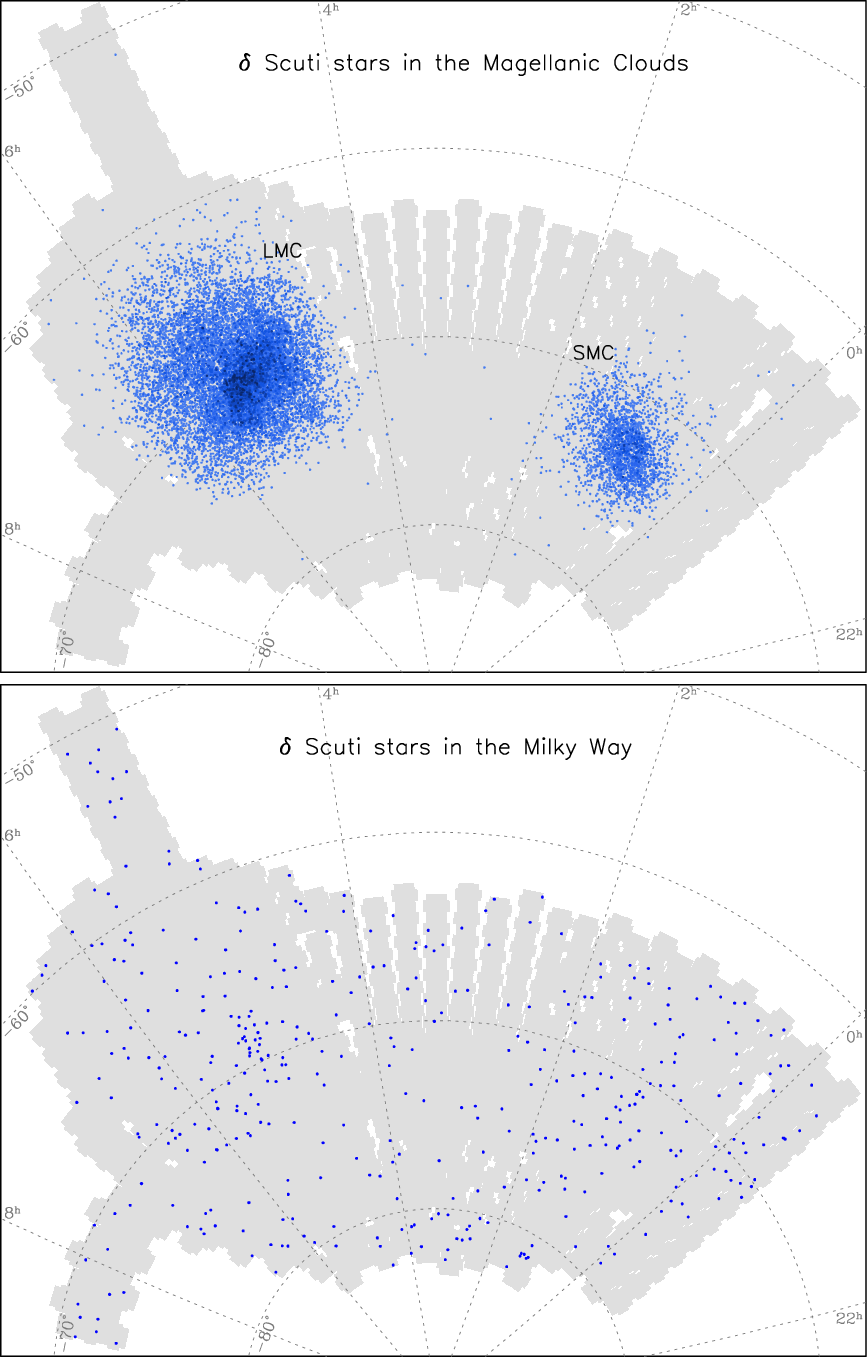}}
\vskip3pt
\FigCap{On-sky distribution of $\delta$~Sct stars in the direction of
  the Magellanic Clouds. {\it Upper panel} displays the positions of
  more than 17\,600 probable members of the LMC and SMC, while {\it
    lower panel} shows $\delta$~Sct variables that are likely part of
  the Milky Way halo. The gray area represents the OGLE footprint in
  the Magellanic System region.}
\end{figure}

\Section{On-sky Map}
The upper panel of Fig.~3 displays the on-sky distribution of about
17\,600 $\delta$~Sct pulsators in the LMC and SMC (Soszy{\'n}ski \etal
2022), while the lower panel shows the positions of approximately 400
SX~Phe variables likely belonging to the Milky Way's halo. The latter
group consists of stars that are at least 1.5~mag brighter than the
average PL relation fulfilled by $\delta$~Sct variables in the LMC or
SMC, respectively.

The spatial distributions of different stellar populations provide
valuable information about their history. For example, the ancient
population of RR~Lyr pulsators (Soszy{\'n}ski \etal 2016) in the LMC
exhibits a structure that can be approximated by a triaxial ellipsoid
without any additional substructures (\eg Jacyszyn-Dobrzeniecka \etal
2017). Conversely, classical Cepheids, which are stars younger than
300 million years, tend to concentrate in the LMC bar and spiral arms
(\eg So\-szy{\'n}ski \etal 2015a, Jacyszyn-Dobrzeniecka \etal
2016). $\delta$~Sct stars also follow the bar and spiral arms of the
LMC (Fig.~3), although this pattern is not as distinct as in the case
of classical Cepheids, indicating that the majority of our
$\delta$~Sct sample consists of intermediate-age stars.

In the region of the LMC bar, the maximum surface density of
$\delta$~Sct stars from our catalog exceeds 600 objects per square
degree. The surface density drops almost to zero at a distance of 5
degrees from the center of the LMC (toward the South) or 9 degrees
(toward the North), while the distribution of RR~Lyr stars extends
to much larger distances (\eg Soszy{\'n}ski \etal 2016). This indicates
that our collection of $\delta$~Sct pulsators in the Magellanic Clouds
is primarily comprised of Population~I stars, while the majority of
Population~II SX~Phe variables in the LMC and SMC are too faint to be
detected in the OGLE frames. Of course, this statement does not apply
to foreground $\delta$~Sct stars in the halo of the Milky Way, which
by definition belong to Population~II.

\Section{Period--Luminosity Relations}
\vskip5pt
The investigation of the PL relations followed by $\delta$~Sct
pulsators is one of the most important applications of our
collection. Many empirical calibrations of the PL relations for
$\delta$~Scuti stars have been reported in the literature (\eg Nemec
\etal 1994, Cohen and Sarajedini 2012, Ziaali \etal 2019, Jayasinghe
\etal 2020, Barac \etal 2022, Ngeow \etal 2023), but most were based
on nearby variables with well-determined parallaxes or SX~Phe stars
identified in the Galactic globular clusters, the distances to which
were known from other standard candles, for example RR~Lyr stars.

\vskip3pt
The LMC has many advantages in the context of studying the
distribution of various classes of pulsating stars in the PL
plane. The close proximity, favorable orientation, and low average
reddening toward this galaxy offers a unique opportunity for
conducting in-depth analyses of its stellar component. The LMC hosts
large and diverse populations of variable stars, including some of the
largest known samples of Cepheids, RR Lyr stars, and long-period
variables. Moreover, the distance to the LMC is currently known to an
unprecedented accuracy of 1\% (Pietrzy{\'n}ski \etal 2019).

\vskip3pt
Previous efforts to measure the PL relationships for $\delta$~Sct
stars in the LMC (McNamara \etal 2007, Garg \etal 2010, Poleski \etal
2010, McNamara 2011) relied on small or biased samples of
variables. Recently, Mart{\'\i}nez-V{\'a}zquez \etal (2022) gathered
data for approximately 4000 extragalactic $\delta$~Sct variables
(primarily from the LMC) and investigated their distribution in the PL
plane. They concluded that extragalactic $\delta$~Sct stars exhibit a
single PL relationship with a notable change in slope occurring at a
period of around 0.09~d.
\begin{figure}[htb]
\includegraphics[width=12.7cm]{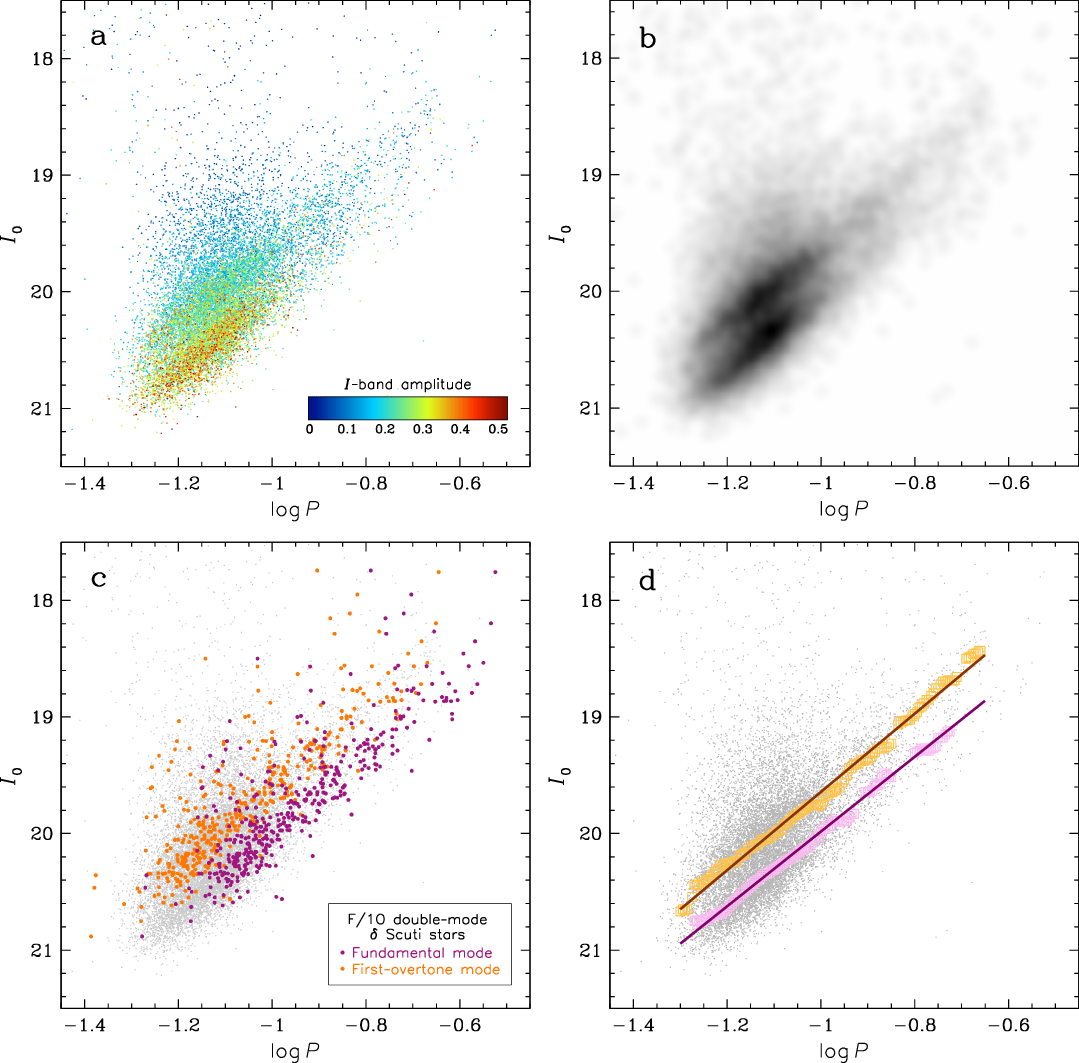}
\FigCap{Extinction-corrected {\it I}-band PL diagram for $\delta$~Sct
  stars in the LMC. {\it Panel~a}: the colors of the points represent
  peak-to-peak amplitudes of the {\it I}-band light curves, as
  indicated by the scale in the bottom right corner. {\it Panel~b}: density
  map of the points in the PL diagram. {\it Panel~c}: PL diagram for
  double-mode $\delta$~Sct stars pulsating in the fundamental (purple
  points) and first-overtone (orange points) modes. Background gray
  dots represent single-mode $\delta$~Sct stars. {\it Panel~d}: linear
  least-square fits to the fundamental-mode (purple line) and
  first-overtone (orange line) PL relations of $\delta$~Sct stars in
  the LMC. Yellow and pink squares indicate local maxima of the
  magnitude distribution.}
\end{figure}

\vskip3pt
The OGLE collection of about 15\,000 $\delta$~Sct stars in the LMC
enables us to verify these results. In Fig.~4, we present four
versions of the extinction-corrected {\it I}-band PL diagram for our
sample. In the upper left panel (a), different colors of points denote
different amplitudes of brightness variations. In this diagram, two
linear PL relationships can be discerned, with variables of larger
amplitudes prevailing along the lower ridge. Due to the significant
dispersion of points around these relationships, in panel~b of Fig.~4,
we provide a density map of the points on this diagram. There is no
doubt that $\delta$~Sct stars in the LMC follow two PL relationships
without apparent changes in slope. This contradicts the findings of
Mart{\'\i}nez-V{\'a}zquez \etal (2022), who reported a single
segmented PL relation, which probably was an illusion resulting from
incompleteness of the prior catalogs of extragalactic $\delta$~Sct
stars.

\vskip3pt
The identification of the pulsation modes corresponding to both PL
ridges can be achieved by plotting double-mode $\delta$~Sct pulsators
with period ratios in the range of 0.755--0.785, \ie corresponding to
the simultaneous oscillations in the fundamental and first-overtone
modes (see Section~9). The color symbols in panel~c of Fig.~4
unambiguously indicate that the lower ridge is populated by
$\delta$~Sct variables pulsating in the fundamental mode, while the
upper ridge is composed of the first-overtone pulsators.

In order to fit the most precise regression lines to both relations in
the {\it I}-band, we constructed histograms of brightness for
consecutive period bins (each with a bin size of 0.05 in $\log{P}$,
moved by $\Delta\log{P}=0.005$). Then, we found two local maxima
(corresponding to the fundamental-mode and the first-overtone ridges)
of the magnitude distributions for each period bin. Any unreliable
determinations of the maxima (\eg due to a limited number of stars
within a bin) were excluded. Finally, we performed linear least-square
fits to the obtained points. The result of our procedure is shown in
panel~d of Fig.~4 and summarized in Table~2. The same method was used
to fit the {\it V}-band PL relations as well as the period--$W_I$ (PW)
relations, where $W_I$ is an extinction-insensitive Wesenheit index,
defined as $W_I=I-1.55(V-I)$.

The dispersion of points around the average PL and PW relations is
significant ($\sigma\approx0.2$~mag), which may stem from measurement
errors of the photometry, blending by unresolved sources, the geometry
of the LMC, differential interstellar extinction, as well as the
diversity of stellar populations present in our sample. In particular,
it is recognized that SX~Phe stars are systematically underluminous
relative to Population~I $\delta$~Sct pulsators (McNamara \etal 2007).

\begin{figure}[t]
\includegraphics[width=12.7cm]{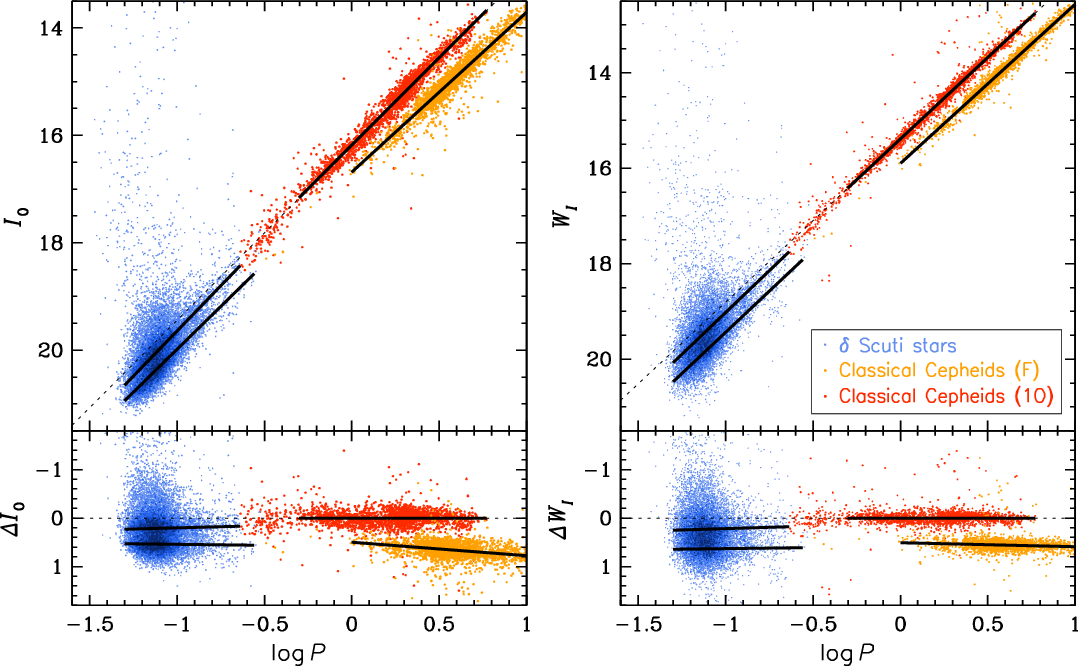}
\FigCap{PL ({\it left panels}) and PW ({\it right panels}) diagrams
  for $\delta$~Sct stars and classical Cepheids in the LMC. Blue,
  orange, and red points mark $\delta$~Sct variables, fundamental-mode
  classical Cepheids, and first-overtone classical Cepheids,
  respectively. Solid lines represent fits to the PL and PW
  relations. Dashed lines are the extensions of the PL and PW
  relationships for first-overtone Cepheids with periods longer than
  0.5~d. {\it Lower panels} show the residuals with respect to the fit
  applied to the first-overtone classical Cepheids with periods longer
  than 0.5~d.}
\end{figure}

\MakeTableee{c@{\hspace{15pt}}
c@{\hspace{15pt}}
c@{\hspace{15pt}}}{12.5cm}
{Period--Luminosity and Period-Wesenheit Relations for $\delta$~Sct stars in the LMC}
{\hline
\noalign{\vskip3pt}
Mode of pulsation & $\alpha$ & $\beta$ \\
\noalign{\vskip3pt}
\hline
\noalign{\vskip3pt}
\multicolumn{3}{c}{$I_0=\alpha\log{P}+\beta$} \\
\noalign{\vskip2pt}
Fundamental    & $-3.203\pm0.024$ & $16.778\pm0.025$ \\
First-overtone & $-3.353\pm0.027$ & $16.290\pm0.026$ \\
\noalign{\vskip3pt}
\hline
\noalign{\vskip3pt}
\multicolumn{3}{c}{$V_0=\alpha\log{P}+\beta$} \\
\noalign{\vskip2pt}
Fundamental    & $-3.048\pm0.025$ & $17.341\pm0.027$ \\
First-overtone & $-3.154\pm0.028$ & $16.822\pm0.029$ \\
\noalign{\vskip3pt}
\hline
\noalign{\vskip3pt}
\multicolumn{3}{c}{$W_I=\alpha\log{P}+\beta$} \\
\noalign{\vskip2pt}
Fundamental    & $-3.449\pm0.026$ & $15.986\pm0.028$ \\
First-overtone & $-3.522\pm0.029$ & $15.501\pm0.030$ \\
\noalign{\vskip3pt}
\hline}

The PL and PW relations for $\delta$~Sct stars (Table~2) are
distinctly steeper than the relations for classical Cepheids
(Soszy{\'n}ski \etal 2015a) pulsating in the same modes. In Fig.~5, we
display the {\it I}-band PL diagram (left panel) and PW diagram (right
panel) for $\delta$~Sct stars and classical Cepheids in the LMC. The
first-overtone variables lay along a continuous ridge in both
diagrams, whereas the fundamental-mode pulsators demonstrate a
discontinuity within the period range of approximately 0.3--1.0~d,
except for several multimode Cepheids with the fundamental-mode
periods falling within this range.

The lower panels of Fig.~5 show the residuals with respect to the
linear fit applied to the first-overtone classical Cepheids with
periods longer than 0.5~d. The dashed line, which is an extension of
this relation, lies above the majority of the $\delta$~Sct stars,
confirming the non-linear nature of the continuous PL and PW
relationships for first-overtone Cepheids and $\delta$~Sct
variables. Moreover, most short-period overtone Cepheids also reside
below this dashed line, indicating that the change in the slope of the
PL and PW relations occurs at period of about 0.5~d
($\log{P}\approx0.3$). This break in the PL relation for
first-overtone classical Cepheids was first noticed by Soszy{\'n}ski \etal
(2008). Recently, Ripepi \etal (2022) confirmed the non-linearity of
the first-overtone PL and PW relations in the near-infrared bands,
pinpointing the break point at $P_{\rm 1O}=0.58\pm0.1$~d.

\Section{Multimode ${\pmb \delta}$~Sct Stars}
Stars that pulsate in multiple modes are attractive targets for
asteroseismological research. Within the general population of
$\delta$~Sct stars, most objects are low-amplitude variables with a
number of non-radial modes simultaneously excited (Breger
2000). However, due to the limitations of the OGLE photometry, the
proportions of low- and high-amplitude variables are inverted in our
collection. The predominant portion of LMC $\delta$~Sct stars within
our sample demonstrates high-amplitude oscillations in the fundamental
or first-overtone modes.

In our collection, we provide up to three pulsation periods per
star. However, for the majority of variables, only a dominant period
could be reliably measured. Secondary or tertiary periods were
identified only when their amplitudes exceeded the detection
thresholds of the OGLE photometry. As a result, the final version of
our catalog includes 621 double-mode and only 18 triple-mode
$\delta$~Sct variables.

\begin{figure}[t]
\includegraphics[width=12.5cm]{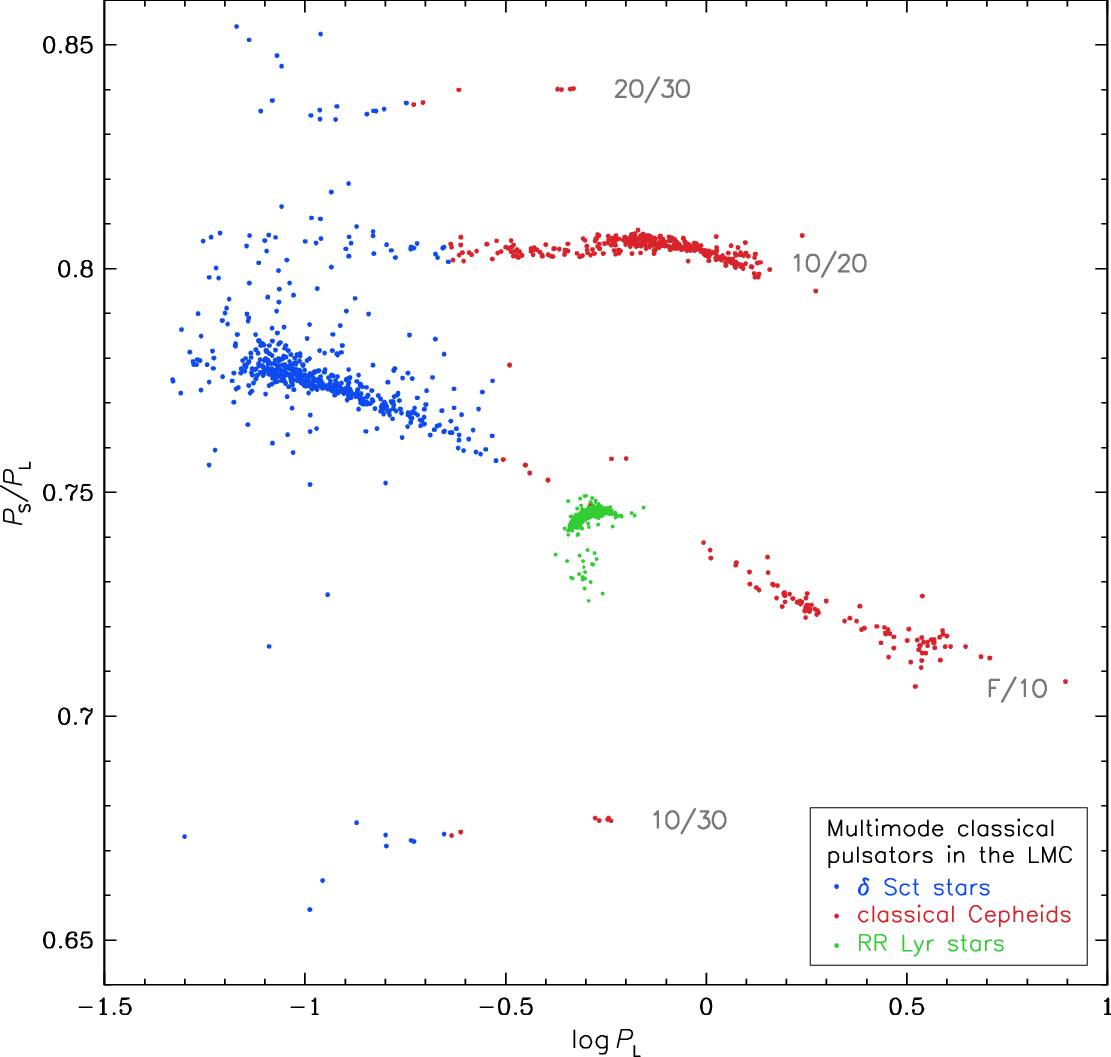}
\FigCap{Petersen diagram for multimode $\delta$~Sct stars (blue
  points), classical Cepheids (red points), and RR~Lyr stars (green
  points) in the LMC.}
\end{figure}

Fig.~6 shows the Petersen diagram (a plot of the ratio between two
periods against the logarithm of the longer one) for multimode
$\delta$~Sct stars, classical Cephe\-ids (Soszy{\'n}ski \etal 2015b), and
RR~Lyr variables (Soszy{\'n}ski \etal 2016) in the LMC. As expected, we
predominantly detected stars oscillating in two or three low-order
radial modes, particularly in the fundamental and first-overtone modes
(F/1O). Roughly two-thirds of all multimode $\delta$~Sct pulsators in
our sample have these two modes excited. Additionally, our dataset
includes about 50 double- and triple-mode $\delta$~Sct stars
simultaneously pulsating in the first, second, or third overtones (in
various configurations), and about 80 variables exhibiting secondary
periods very close to the primary ones. The latter phenomenon may
indicate the presence of the non-radial modes.

Fig.~6 vividly illustrates the continuity between $\delta$~Sct stars
and classical Cephe\-ids in the Petersen diagram. The choice of
pulsation periods that distinguishes both classes of variable stars is
a matter of convention. In the OCVS, we adopted $P_{\rm F}=0.3$~d for
the fundamental mode, $P_{\rm 1O}=0.23$~d for the first-overtone, and
$P_{\rm 2O}=0.185$~d for the second overtone.

The Petersen diagram is a powerful tool to constrain stellar
parameters such as masses or metallicities of multimode pulsators (\eg
Petersen and Christensen-Dalsgaard 1996, Poretti \etal 2005, Netzel
\etal 2022). Fig.~7 shows a zoom-in of the Petersen diagram focusing
on the region occupied by F/1O $\delta$~Sct stars in the Milky Way
(Soszy{\'n}ski \etal 2021), SMC (Soszy{\'n}ski \etal 2022), and LMC (this
work).

\begin{figure}[h]
\centerline{\includegraphics[width=8.0cm]{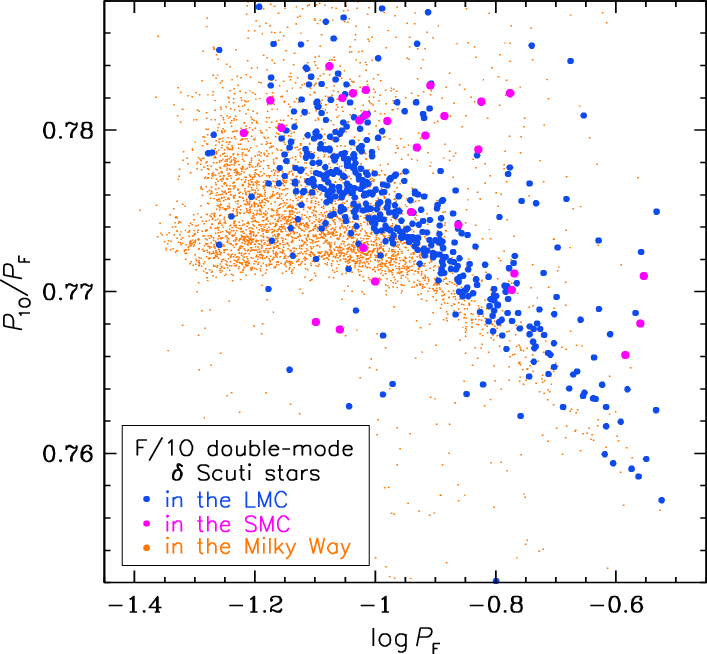}}
\FigCap{Petersen diagram for F/1O double-mode $\delta$~Sct stars in
  the LMC (blue points), SMC (magenta points), and Milky Way (orange
  points).}
\end{figure}

The Galactic double-mode pulsators exhibit a characteristic splitting
of the period--period ratio sequence at the short-period end. This
reflects the division of $\delta$~Sct variables into the Population~I
stars ($P_{\rm 1O}/P_{\rm F}\approx0.773$) and SX~Phe stars
($P_{\rm 1O}/P_{\rm F}\approx0.778$, Breger 2000). Double-mode
pulsators with such short periods are not present in the OGLE
collection of $\delta$~Sct stars in the Magellanic Clouds due to a
selection bias. These stars are too faint to be identified through
OGLE photometry. Nonetheless, it is evident that the period--period
ratio sequence for $\delta$~Sct stars in the LMC is situated above the
sequence for Galactic stars, and even further above lie the SMC
pulsators. Double-mode F/1O classical Cepheids in the Milky Way, LMC,
and SMC exhibit analogous behavior (\eg Udalski \etal 2018), which can
be attributed to the different metallicities among these three
galaxies.

\Section{${\pmb \delta}$~Sct Stars in Binary Systems}
\begin{figure}[p]
\centerline{\includegraphics[width=13.3cm]{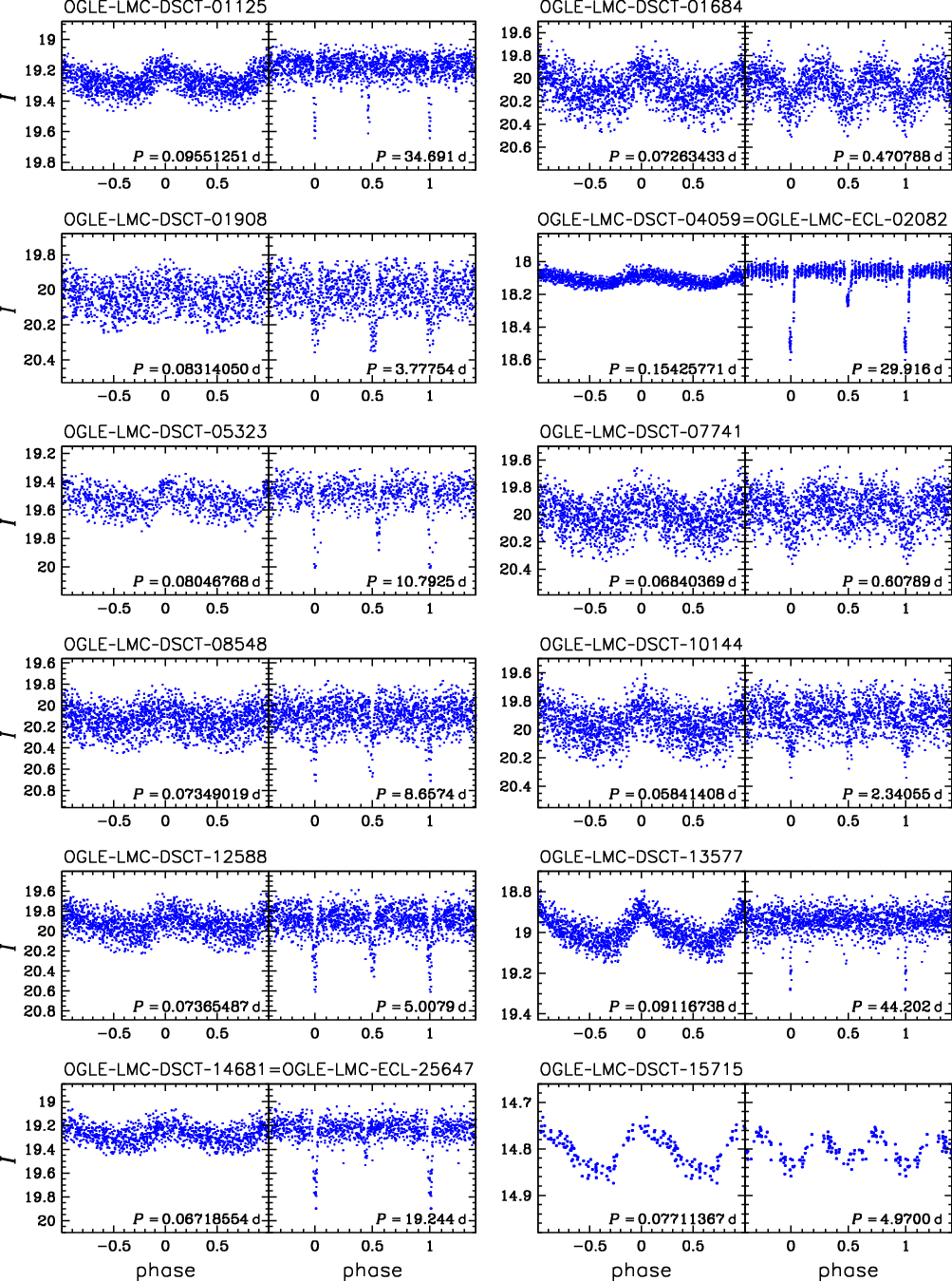}}
\vspace*{2pt}
\FigCap{Disentangled {\it I}-band light curves of $\delta$~Sct stars
  showing additional eclipsing or ellipsoidal modulation. In each
  pair, {\it left panel} displays the pulsation light curve, while
  {\it right panel} shows the eclipsing/ellipsoidal light curve after
  subtracting the pulsation component.}
\end{figure}

Eclipsing binary systems offer a unique opportunity to accurately
measure the physical parameters of the stellar components, such as
masses, radii, luminosities, and temperatures. As a result, binaries
comprising pulsating stars serve as excellent testbeds for
asteroseismology and stellar evolution theory. $\delta$~Sct variables
within binary systems are not infrequently encountered
entities. According to the updated version of the Liakos and Niarchos
(2017)
catalog\footnote{\it https://alexiosliakos.weebly.com/catalogue.html}, a
total of 367 binaries with a $\delta$~Sct component have been known so
far, including 34 such objects identified by the OGLE team
(Pietrukowicz \etal 2020, Soszy{\'n}ski \etal 2021). However, all of these
systems are situated within the Milky Way.

While searching for potential secondary periods in the light curves of
$\delta$~Sct stars in the LMC, we discovered 12 objects demonstrating
additional variability caused by binarity, \ie eclipses or ellipsoidal
variations. The details of these stars, including their coordinates,
mean magnitudes, pulsation and orbital periods, are provided in
Table~3. Fig.~8 displays their disentangled pulsating and
eclipsing/ellipsoidal light curves. To our best knowledge, these stars
are the first known extragalactic candidates for binary systems
containing $\delta$~Sct components. The spectroscopic examination of
these objects will be challenging due to their low apparent
luminosity. Nevertheless, the long-term OGLE photometry could be
useful in studying \eg the stability of their orbital periods or the
apsidal motion in the systems featuring eccentric orbits.

\MakeTableee{l@{\hspace{6pt}}
c@{\hspace{5pt}}
c@{\hspace{6pt}}
c@{\hspace{6pt}}
c@{\hspace{6pt}}
c@{\hspace{4pt}}
l@{\hspace{0pt}}}
{12.5cm}{$\delta$~Sct stars with additional eclipsing or ellipsoidal modulation}
{\hline
\noalign{\vskip3pt}
\multicolumn{1}{c}{Identifier}
& R.A.
& Dec.
& $\langle{I}\rangle$
& $\langle{V}\rangle$
& $P_{\rm puls}$
& \multicolumn{1}{c}{$P_{\rm orb}$} \\
& [J2000.0]
& [J2000.0]
& [mag]
& [mag]
& [d]
& \multicolumn{1}{c}{[d]} \\
\noalign{\vskip3pt}
\hline
\noalign{\vskip3pt}
OGLE-LMC-DSCT-01125 & 05\uph15\upm35\zdot\ups13 & $-68\arcd36\arcm55\zdot\arcs3$ & 19.162 & 19.664 & 0.09551251 & 34.691 \\
OGLE-LMC-DSCT-01684 & 05\uph26\upm57\zdot\ups48 & $-70\arcd05\arcm24\zdot\arcs1$ & 20.053 & 20.447 & 0.07263433 & ~~0.47079 \\
OGLE-LMC-DSCT-01908 & 05\uph30\upm09\zdot\ups29 & $-69\arcd02\arcm46\zdot\arcs1$ & 20.004 & 20.366 & 0.08314050 & ~~3.77754 \\
OGLE-LMC-DSCT-04059 & 04\uph52\upm56\zdot\ups49 & $-69\arcd32\arcm20\zdot\arcs0$ & 18.057 & 18.905 & 0.15425771 & 29.916 \\
OGLE-LMC-DSCT-05323 & 05\uph02\upm24\zdot\ups90 & $-67\arcd09\arcm49\zdot\arcs1$ & 19.480 & 20.094 & 0.08046768 & 10.7925 \\
OGLE-LMC-DSCT-07741 & 05\uph16\upm46\zdot\ups31 & $-69\arcd11\arcm19\zdot\arcs5$ & 19.939 & 20.250 & 0.06840369 & ~~0.60789 \\
OGLE-LMC-DSCT-08548 & 05\uph20\upm30\zdot\ups89 & $-68\arcd56\arcm09\zdot\arcs4$ & 20.066 & 20.597 & 0.07349019 & ~~8.6574 \\
OGLE-LMC-DSCT-10144 & 05\uph27\upm54\zdot\ups03 & $-70\arcd10\arcm55\zdot\arcs0$ & 19.891 & 20.322 & 0.05841408 & ~~2.34055 \\
OGLE-LMC-DSCT-12588 & 05\uph40\upm47\zdot\ups14 & $-68\arcd15\arcm25\zdot\arcs0$ & 19.860 & 20.324 & 0.07365487 & ~~5.0079 \\
OGLE-LMC-DSCT-13577 & 05\uph49\upm02\zdot\ups91 & $-67\arcd37\arcm37\zdot\arcs3$ & 18.935 & 19.616 & 0.09116738 & 44.202 \\
OGLE-LMC-DSCT-14681 & 06\uph01\upm37\zdot\ups06 & $-70\arcd37\arcm30\zdot\arcs6$ & 19.219 & 19.962 & 0.06718554 & 19.244 \\
OGLE-LMC-DSCT-15715 & 06\uph50\upm22\zdot\ups94 & $-79\arcd20\arcm23\zdot\arcs8$ & 14.803 & 15.674 & 0.07711367 & ~~4.9700 \\
\noalign{\vskip3pt}
\hline}

\Section{Summary}
We presented the OGLE collection of about 15\,000 $\delta$~Sct
variables in the LMC. Approximately, two-thirds of these stars
represent new discoveries. This compilation constitutes the most
extensive sample of extragalactic $\delta$~Sct stars published to
date. Our catalog is a part of the OCVS, which presently comprises
around 1.1 million manually selected and classified variable stars in
the Milky Way and the Magellanic Clouds. The OCVS facilitates
comparative research on pulsating stars in different stellar
environments. The extensive and well-sampled OGLE light curves in the
standard photometric system offer opportunities to investigate exotic
modes in pulsating stars, assess the stability of periods, and detect
pulsating stars in binary systems.

This paper provides just a glimpse of the potential research that can
be carried out on the variables within our collection. We presented
the on-sky distribution of $\delta$~Sct stars in the Magellanic
System, derived empirical PL relations for fundamental-mode and
first-overtone pulsators and compared them to the PL relations for
classical Cepheids. Additionally, we conducted a comparison of period
ratios in multimode $\delta$~Sct variables originated from the Milky
Way and the Magellanic Clouds. Finally, we reported the discovery of
the first-known candidates for extragalactic eclipsing binaries
containing a $\delta$~Sct component.

\Acknow{This work has been funded by the National Science Centre,
  Poland, grant no.~2022/45/B/ST9/00243. For the purpose of Open
  Access, the author has applied a CC-BY public copyright license to
  any Author Accepted Manuscript (AAM) version arising from this
  submission.}

\end{document}